\documentclass{ws-procs10x7-mod}

\newcommand{\compact}[1]{}    

\usepackage{amsmath}
\usepackage{url}
\newcommand{\as}{\alpha_s}
\newcommand{\ifb}{\,\mathrm{fb}^{-1}}
\newcommand{\GeV}{\,\mathrm{GeV}}
\newcommand{\jet}{\,\mathrm{jet}}
\newcommand{\jets}{\,\mathrm{jets}}
\newcommand{\bra}{\langle}
\newcommand{\ket}{\rangle}
\newcommand{\eg}{e.g.\ }
\newcommand{\ie}{i.e.\ }
\newcommand{\cN}{{\cal N}}
\begin{document}


\title{Developments in perturbative QCD}

\author{Gavin P. Salam}

\address{LPTHE, Universities of Paris VI and VII and CNRS UMR 7589, Paris,
  France.}

\twocolumn[\maketitle\abstract{ A brief review of key recent
  developments and ongoing projects in perturbative QCD theory, with
  emphasis on conceptual advances that have the potential for impact
  on LHC studies.  Topics covered include: twistors and new recursive
  calculational techniques; automation of one-loop predictions;
  developments concerning NNLO calculations; the status of Monte Carlo
  event generators and progress in matching to fixed order; analytical
  resummation including the push to NNLL, automation and gap between
  jets processes; and progress in the understanding of saturation at
  small~$x$.}]

\section{Introduction}



A significant part of today's research in QCD aims to provide tools to
help better constrain the standard model and find what may lie beyond
it.
%
%
For example one wishes to determine, as accurately as possible, the
fundamentals of the QCD and electroweak theories, such as $\as$, quark
masses, and the elements of the CKM matrix.  One also needs precise
information about `pseudo-fundamentals,' quantities such as parton
distribution functions (PDFs) that could be predicted if we knew how
to solve non-perturbative QCD, but which currently must be deduced
from experimental data.  Finally, one puts this information together
to predict the QCD aspects of both backgrounds and signals at high
energy colliders, particularly at the Tevatron and LHC, to help
maximise the chance of discovering and understanding any new physics.

Other facets of QCD research seek to extend the boundaries of our
knowledge of QCD itself. The underlying Yang-Mills field theory is
rich in its own right, and unexpected new perturbative structures have
emerged in the past two years from considerations of string theory.  In the
high-energy limit of QCD it is believed that a new state appears, the
widely studied colour glass condensate, which still remains to be well
understood. And perhaps the most challenging problem of QCD is that of
how to relate the partonic and hadronic degrees of freedom.

Given the practical importance of QCD for the upcoming LHC programme,
this talk will concentrate on results (mostly since the 2003 Lepton-Photon
symposium) that bring us closer to the well-defined goals mentioned in
the first paragraph. Some of the more explorative aspects will also be
encountered as we go along, and one should remember that there is
constant cross-talk between the two. For example: improved
understanding of field theory helps us make better predictions for
multi-jet events, which are important backgrounds to new physics; and
by comparing data to accurate perturbative predictions one can attempt
to isolate and better understand the parton-hadron interface.

The first part of this writeup will be devoted to results at fixed
order. At tree level we will examine new calculational methods that
are much more efficient than Feynman graphs; we will then consider NLO
and NNLO calculations and look at the issues that arise in going from
the Feynman graphs to useful predictions. 

One of the main uses of fixed-order order predictions is for
understanding rare events, those with extra jets.  In the second part
of the writeup we shall instead turn to resummations, which help us
understand the properties of typical events. 

Throughout, the emphasis will be on the conceptual advances rather
than the detailed phenomenology. Due to lack of space, some active
current topics will not be covered, in particular exclusive
QCD. Others are discussed elsewhere in these proceedings.\cite{Others}

\section{Fixed-order calculations}

\subsection{Tree-level amplitudes and twistors}

Many searches for new physics involve signatures with a large number
of final-state jets. Even for as basic a process as $t\bar t$
production, the most common decay channel (branching ratio of 46\%),
$t\bar t \to b \bar b W^{+} W^{-} \to b\bar b q q \bar q \bar q$
involves 6 final-state jets, to which there are large QCD multi-jet
backgrounds.\cite{JusteThisConf} And at the LHC, with $10\ifb$ (1
year) of data, one expects of the order of 2000 events with 8 or more
jets\cite{Draggiotis:2002hm} ($p_t(\mathsf{jet}) > 60 \GeV$,
$\theta_{ij} > 30\deg$, $|y_{i}| < 3$).

For configurations with such large numbers of jets, even tree-level
calculations become a challenge --- for instance $gg \to 8g$ involves
10525900 Feynman diagrams (see ref.\cite{GloverCAPPTalk}). In the
1980's, techniques were developed to reduce the complexity of such
calculations.\cite{Dixon:1996wi} Among them
colour decomposition,\cite{ColourOrdering} where one separates the
colour and Lorentz structure of the amplitude,
\begin{multline}
  {\cal A}^\mathsf{tree}(1,2,\ldots,n) = g^{n-2} \sum_{perms}  \\
  \underbrace{
    \mathsf{Tr} (T_{1} T_{2} \ldots
    T_{n})}_{\textsf{colour struct.}}
  \underbrace{
    A^\mathsf{tree}(1,2,\ldots,n)}_{
    \textsf{colour ordered amp.}
  }\,;
\end{multline}
the use\cite{SpinorFormalism} of spinor products $\bra i j \ket \equiv
\bra i^{-} | j^{+}\ket = \overline{u_-(k_i)}u_+(k_j)$ and $[ij] = \bra
i^{+} | j^{-}\ket$ as the key building blocks for writing amplitudes;
and the discovery\cite{ParkeTaylor} and subsequent
proof\cite{BerendsGieleRecursion} of simple expressions for the subset
of amplitudes involving the maximal number of same-helicity spinors
(maximum helicity violating or MHV amplitudes), \ie $n-2$ positive
helicity spinors for an $n$-gluon amplitude:
\begin{equation}
  A^{\mathrm{tree}}( - - + +\ldots) = 
  \frac{i\bra 12\ket^4}{\bra12\ket
    \bra23\ket \ldots \bra n1\ket}\,.
\end{equation}
Ref.\cite{BerendsGieleRecursion} also provided a computationally
efficient recursion relation for calculating amplitudes with arbitrary
numbers of legs, and with these and further
techniques,\cite{Caravaglios:1995cd} numerous programs (\eg
MadEvent,\cite{Maltoni:2002qb} ALPGEN,\cite{Alpgen}
HELAC/PHEGAS,\cite{Kanaki:2000ms} CompHEP,\cite{Pukhov:1999gg}
GRACE,\cite{Yuasa:1999rg} Amegic\cite{Krauss:2001iv}) are able to
provide results for processes with up to 10 legs.

The past two years have seen substantial unexpected progress in the
understanding of multi-leg tree-level processes. It was initiated by
the observation (first made by Nair\cite{Nair:1988bq}) that helicity
amplitudes have a particularly simple form in `twistor' space, a space
where a Fourier transform has been carried out with respect to just
positive helicity spinors. In twistor space a duality
appears\cite{Witten:2003nn} between the weakly-coupled regimes of a
topological string theory and $\cN=4$ SUSY Yang-Mills. This has led to the
postulation, by Cachazo, Svrcek and Witten (CSW),\cite{CSW} of rules
for deriving non-MHV $\cN=4$ SUSY amplitudes from MHV ones,
illustrated in fig.~\ref{fig:CSW-tules}.  For purely gluonic
amplitudes the results are identical to plain QCD,\cite{TreeIsSUSY}
because tree level SUSY amplitudes whose external legs are gluons have
only gluonic propagators.
\begin{figure}[htbp]
  \centering
  \includegraphics[width=0.45\textwidth]{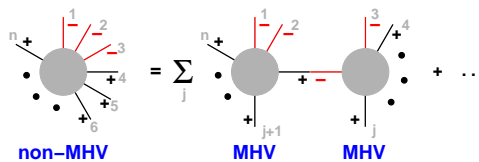}
  \caption{Graphical illustration of the CSW rules: by joining
    together two MHV amplitudes with an off-shell scalar propagator
    one obtains an amplitude with an extra negative helicity (NMHV).}
  \label{fig:CSW-tules}
\end{figure}

Recently a perhaps even more powerful set of recursion relations was
proposed by Britto, Cachazo and Feng (BCF),\cite{BCF} which allows one
to build a general $n$-leg diagram by joining together pairs of
on-shell sub-diagrams. This is made possible by continuing a pair of
reference momenta into the complex plane. The proof\cite{BCFW} of
these relations (and subsequently also of the CSW
rules\cite{Risager:2005vk}) is remarkably simple, based just on the
planar nature of colour-ordered tree diagrams, their analyticity
structure, and the asymptotic behaviour of known MHV amplitudes.

The discovery of the CSW and BCF recursion relations has spurred
intense activity, about 150 articles citing the original
papers\cite{Witten:2003nn,CSW} having appeared in the 18 months
following their publication. Questions
addressed include the derivation of simple expressions for specific
amplitudes,\cite{Kosower:2004yz}
the search for computationally efficient recursive
formulations,\cite{Bena:2004ry} extensions to processes with fermions
and gluinos,\cite{Wu:2004fb}
Higgs bosons,\cite{Dixon:2004za}
electroweak bosons,\cite{Bern:2004ba} gravity,\cite{Bedford:2005yy}
and the study of multi-gluon collinear
limits.\cite{Birthwright:2005vi} This list is necessarily incomplete
and further references can be found in a recent
review\cite{Cachazo:2005ga} as well as below, when we discuss
applications to loop amplitudes.

\subsection{One-loop amplitudes}

For quantitatively reliable predictions of a given process it is
necessary for it to have been calculated to next-to-leading order
(NLO). A wide variety of NLO calculations exists, usually in the form
of publicly available programs,\cite{hepcode} that allow one to make
predictions for arbitrary observables within a given process.  The
broadest of these programs are the MCFM,\cite{MCFM}
NLOJET\cite{NLOJET} and PHOX\cite{PHOX} families.

For a $2\to n$ process the NLO calculation involves the $2\to n+1$
tree-level diagram, the $2\to n$ 1-loop diagram, and some method for
combining the tree-level matrix element with the loop contributions,
so as to cancel the infrared and collinear divergences present in both
with opposite signs. We have seen above that tree-level calculations
are well understood, and dipole subtraction\cite{dipole} provides a
general prescription for combining them with the corrresponding 1-loop
contributions.
The bottleneck in such calculations remains the determination of the
1-loop contribution.  Currently, $2\to 3$ processes are feasible,
though still difficult, while as yet no full $2\to 4$ 1-loop QCD
calculation has been completed.

In view of the difficulty of these loop calculations, a welcome
development has been the compilation, by theorists and experimenters
at the Les Houches 2005 workshop, Physics at TeV
Colliders,\cite{LesHouches} of a realistic prioritised wish-list of
processes.  Among the most interesting remaining $2\to 3$ processes
one has $pp\to WW+\mathrm{jet}$, $pp\to VVV$ ($V=W$ or $Z$) and $pp
\to H + 2\jets$.  The latter can be considered a `background' to Higgs
production via vector boson fusion, insofar as the isolation of the
vector-boson fusion channel for Higgs production would allow
relatively accurate measurements of the Higgs
couplings.\cite{HiggsCouplings} A number of $2\to 4$ processes are
listed as backgrounds to $t\bar t H$ production ($pp\to t\bar tq\bar
q, t\bar tb\bar b$), $WW\to H\to WW$ or to general new physics ($pp\to
V + 3\jets$) or specifically SUSY ($pp\to VVV + \jet$).

Two broad classes of techniques have been used in the past for 1-loop
calculations: those based directly on the evaluation of the Feynman
diagrams (sometimes for the 1-loop-tree
interference\cite{Campbell:1997tv}) and those based on unitarity
techniques to sew together tree diagrams (see the
review\cite{Bern:1996je}). Both approaches are still being actively
pursued.

Today's direct evaluations of 1-loop contributions have, as a starting
point, the automated generation of the full set of Feynman diagrams,
using tools such as QGRAF\cite{QGRAF} and FeynArts.\cite{FeynArts} The
results can be expressed in terms of sums of products of
group-theoretic (\eg colour) factors and tensor one-loop integrals,
such as
\begin{equation}
  \label{eq:1}
  I_{n;\mu_1..\mu_i} = \int d^{4+2\epsilon}\ell \frac{\ell_{\mu_1}\ldots
    \ell_{\mu_i}}{(\ell+k_1)^2\ldots(\ell+k_n)^2}
\end{equation}
Reduction procedures exist (e.g.\cite{TensorReductions1,Davydychev:1991va})
that can be applied recursively so as to express the
$I_{n;\mu_1..\mu_i}$ in terms of known scalar integrals. Such
techniques form the basis of recent proposals for automating the
evaluation of the integrals, where the recursion relations are solved
by a combination of analytical and numerical
methods,\cite{Binoth:2005ff} or purely
numerically,\cite{Giele:2004iy,delAguila:2004nf,vanHameren:2005ed} in
some cases with special care as regards divergences that appear in the
coefficients of individual terms of the recursion relation but vanish
in the sum.\cite{delAguila:2004nf,Giele:2004ub} 
%
Results from these approaches include a new compact form for $gg\to
g\gamma\gamma$ at 1 loop\cite{Binoth:2005ua} and the 1-loop
contribution to the `priority' $pp \to H + 2\jets$
process.\cite{EGZ-Higgs}
Related automated approaches have also been developed in the context
of electroweak
calculations,\cite{Ferroglia:2002mz}
where recently a first $2\to4$ 1-loop result ($e^+e^- \to 4$~fermions)
was obtained.\cite{Denner:2005es}

%
%
%
%
%
%

Another approach\cite{Nagy:2003qn} proposes subtraction terms for
arbitrary 1-loop graphs, such that the remaining part of the
loop integral can carried out numerically in 4 dimensions. The subtraction
terms themselves can be integrated analytically for the sum over graphs
and reproduce all infrared, collinear and ultraviolet divergences.

The above methods are all subject to the problem of the rapidly
increasing number of graphs for multi-leg processes. On the other
hand, procedures based on `sewing' together tree graphs to obtain loop
graphs\cite{Bern:1996je} (cut constructibility approach) can
potentially benefit from the simplifications that emerge from twistor
developments for tree graphs. This works best for $\cN=4$ SUSY QCD,
where cancellations between scalars, fermions and vector particles
makes the `sewing' procedure simplest and for example all gluonic (and
some scalar and gluino) NMHV 1-loop helicity amplitudes are now
known;\cite{Bern:2004ky}
also, conjectures for $\cN=4$ SUSY $n$-leg MHV planar graphs, at any
number of loops, based on 4-gluon two and three-loop
calculations,\cite{AllLoopConjectures} have now been explicitly verified for 5
and 6 gluon two-loop amplitudes.\cite{Buchbinder:2005wp}
%
For $\cN=1$ SUSY QCD, known results
for all MHV amplitudes\cite{Bern:1994cg} have been
reproduced\cite{CSWMHV1loop} and new results exist for some all-$n$
NMHV graphs\cite{Bidder:2005ri} as well as full results for 6
gluons.\cite{Britto:2005ha} For plain QCD, progress has been slower,
though all finite 1-loop graphs ($++++\ldots$, $-+++\ldots$) were
recently presented\cite{Bern:2005ji} and understanding has also been
achieved for divergent graphs,\cite{Brandhuber:2005jw}
including the full result for all 1-loop $(--+++\ldots)$ MHV
graphs.\cite{Forde:2005hh} The prospects for the twistor-inspired
approach are promising and one can hope that it will soon become
practically competitive with the direct evaluation methods.

\subsection{NNLO jet calculations}

NNLO predictions are of interest for many reasons: in those processes
where the perturbative series has good convergence they can help bring
perturbative QCD predictions to the percent accuracy level. In cases
where there are signs of poor convergence at NLO they will hopefully
improve the robustness of predictions and, in all cases, give
indications of the reliability of the series expansion. Finally they
may provide insight in the discussion of the relative importance of
hadronisation and higher-order perturbative
corrections.\cite{DW,GG,Abdallah:2003xz}

So far, NNLO predictions are available mostly only for processes with
3 external legs, such as the total cross section for $Z\to
\mathrm{hadrons}$ or the inclusive $pp\to
W,Z$\cite{Hamberg:1990np,Harlander:2002wh,Ravindran:2003gi} or
Higgs\cite{Harlander:2002wh,Anastasiou:2002yz,Ravindran:2003um} cross
sections. A full list is given in table~1 of Stirling's ICHEP
writeup.\cite{Stirling:2004db}

Most current effort is being directed to the $e^+e^- \to 3\jets$
process, where NLO corrections are often large, and where one is free
of complications from incoming coloured particles. The ingredients
that are needed are the squared 5-parton tree level ($M_{5}$) and
3-parton 1-loop ($M_{3a}$) amplitudes, and the interference between
4-parton tree and 1-loop ($M_{4}$), and between 3-parton tree and
2-loop amplitudes ($M_{3b}$), all of which are now known (for
references see introduction of\cite{Gehrmann-DeRidder:2005cm}). A full
NNLO prediction adds the integrals over phase-space of these
contributions, multiplied by some jet-observable function $J$ that
depends on the momenta $p_i$
\begin{multline*}
  J_{NNLO} = \int d^D\!\Phi_5 M_5 J(p_{1..5}) \,+ \\
  \int d^D\!\Phi_4 M_4 J(p_{1..4}) + \int d^D\!\Phi_3 (M_{3a+b})
  J(p_{1..3})\,.
\end{multline*}
Each of the terms is infrared and collinear (IRC) divergent, because
of the phase space integration ($M_{4,5}$) and/or the loop integral
($M_{3,4}$).  The current bottleneck for such calculations is in
cancelling these divergences for an arbitrary (IRC safe) $J$.

A standard approach at NLO is to introduce subtraction terms (\eg in
the dipole\cite{dipole} formalism), schematically,
\begin{multline*}
  J_{NLO}^{4-jet} = \int d^D\!\Phi_5 (M_5 J(p_{1..5}) - S_5 J(\tilde
  p_{1..4})) \\
  + \int d^D\!\Phi_4 (M_4 J(p_{1..4}) + S_4 J(p_{1..4}))
\end{multline*}
such that each integral is separately finite and that, alone, the
$S_5$ and $S_4$ terms integrate to equal and opposite divergent
contributions (they both multiply the same 4-parton jet
function $J( p_{1..4})$ and the $\tilde p_{1..4}$ are a
specifically designed function of the $p_{1..5}$).  At NNLO a similar
method can be envisaged, and considerable work has gone towards
developing a general
formalism.\cite{Kosower:2002su,Weinzierl:2003fx,Kilgore:2004ty,Frixione:2004is,Somogyi:2005xz,Gehrmann-DeRidder:2005cm}

An alternative approach, sector
decomposition,\cite{Binoth:2000ps,Heinrich:2002rc,Anastasiou:2003gr,Binoth:2004jv}
introduces special distributions $f_{-i}$ (involving plus-functions,
like those in splitting functions), which isolate the
$\epsilon^{-i}$ divergent piece of a given integral
($\epsilon=(D-4)/2$), {\small
\begin{multline*}
  \int d^D\!\Phi_5 M_5 J(p_{1..5}) = 
   \frac1{\epsilon^{4}}\!\!\int d^4\!\Phi_5 f_{-4} M_5 J(p_{1..5})
   \\  + \frac1{\epsilon^{3}}\!\!\int d^4\!\Phi_5 f_{-3} M_5
   J(p_{1..5}) + \ldots\,,
\end{multline*}}%
where the integration is performed in transformed variables that
simplify the separation of divergences. In such an approach one thus
obtains separate results for each power of $\epsilon$ in both real and
virtual terms, making it easy to combine them.

A useful testing ground for a number of these approaches has been
$e^+e^-\to2\jets$.\cite{Binoth:2004jv,Anastasiou:2004qd,Frixione:2004is}
Fundamentally new results are the differential distributions for
$W,Z,H$ production\cite{Anastasiou:2003ds} in the
sector-decomposition approach and the $(\as C_F/2\pi)^3$ contribution
to $\bra 1 - \mathrm{Thrust}\ket$ in $e^+e^- \to 3\jets$, $-20.4\pm4$,
in the `antenna' subtraction approach.\cite{Gehrmann-DeRidder:2005cm}

\subsection{NNLO splitting functions}

\emph{The} landmark calculation of 2004 was probably that of the NNLO
splitting functions by Moch, Vermaseren and Vogt\cite{Moch:2004pa}
(MVV).  These are important for accurate DGLAP evolution of the parton
distributions, as extracted from fixed target, HERA and Tevatron data,
up to LHC scales.

The results for the splitting functions take about 15 pages to
express, though the authors have also provided compact
approximations for practical use. These are gradually being adopted in
NNLO fits.\cite{Martin:2004dh,Alekhin:2005gq} Mostly the NNLO
splitting functions are quite similar to the estimates obtained a few
years ago\cite{NNLOapprox} based on a subset of the moments and known
asymptotic limits. In particular it remains true that the NNLO
corrections are in general small, both compared to NLO and in absolute
terms. The worst region is that of small $x$ for the singlet
distributions, shown in fig.~\ref{fig:dqS}, where the NLO corrections
were large and there is a significant NNLO modification as well.
Studies of small-$x$ resummation suggest that further high-order
effects should be modest,\cite{Ciafaloni:2003rd,Altarelli:2004dq}
though so far only the gluon sector has been studied in detail.

\begin{figure}[htbp]
  \centering
  \includegraphics[width=0.8\columnwidth,height=0.9\columnwidth]{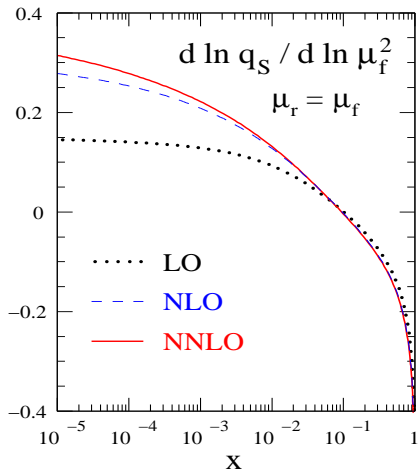}
  \caption{Impact\protect\cite{Moch:2004pa} of NNLO DGLAP corrections
    on the derivative of a toy singlet quark distribution $q_S =
    \sum_i (q_i + \bar q_i)$.}
  \label{fig:dqS}
\end{figure}

Another potentially dangerous region is that of large $x$: the
splitting functions converge well, but the coefficient functions have
$[\as^n \ln^{2n-1} (1-x)]/(1-x)$ enhancements. There are suggestions
that the all-order inclusion of these enhancements via a threshold
resummation may help improve the accuracy of PDF
determinations.\cite{Corcella:2005us} Fresh results from the MVV
group,\cite{Moch:2005ba} for the third order electromagnetic
coefficient functions, quark and gluon form factors, and new threshold
resummation coefficients should provide the necessary elements for yet
higher accuracy at large $x$.  Together with the large-$x$ part of the
NNLO splitting functions these have been used as inputs to the N$^3$LO
soft-collinear enhanced terms for the Drell-Yan and Higgs cross
sections.\cite{Moch:2005ky,Laenen:2005uz}

Finally, as splitting-function calculations approach accuracies the
$1\%$ level, one should consider also the relevance of QED
corrections.\cite{Martin:2004dh,Ward:2005ir}


\subsection{Other accuracy-related issues}

In view of the efforts being devoted to improving the accuracy of
theoretical predictions, it is disheartening to discover that there
are still situations where accuracy is needlessly squandered through
incorrect data-theory comparisons. This is the case for the inclusive
jet cross sections in the cone algorithm at the Tevatron, where a
parameter $R_{sep}=1.3$ is introduced in the cone algorithm used for
the NLO calculation (see e.g.\cite{Wobisch:2004ru}), but not in the
cone algorithm applied to the data.  $R_{sep}$ is the multiple
of the cone radius beyond which a pair of partons is not recombined.
It dates to early theoretical work\cite{Ellis:1992qq} on NLO
corrections to the cone algorithm in hadron collisions, when there was
no public information on the exact jet algorithm used by the
experiments: $R_{sep}$ was introduced to parametrise that ignorance.

The use of $R_{sep}$ in just the theory introduces a spurious NLO
correction (at the $5-10\%$ level\cite{Ellis:1992qq}), meaning that
the data-theory comparisons are only good to LO. The size of the
discrepancy is comparable to the NLO theory uncertainty. As this is
smaller than experimental errors, for now the practical impact is limited.
However, as accuracies improve it is essential that theory-experiment
comparisons be done consistently, be it with properly used cone
algorithms\cite{RunII-jet-physics} or with the (more straightforward
and powerful) $k_t$ algorithm,\cite{JetratesHH} which is finally
starting to be investigated.\cite{KtTev}

An accuracy issue that is easily overlooked when discussing QCD
developments is the non-negligible impact of electroweak effects at
large scales. The subject has mostly been investigated for leptonic
initial and final states at a linear collider (ILC).
The dominant contributions go as $\alpha_{EW} \ln^2 P_t/M_W$. Since
the LHC can reach transverse momenta ($P_t$) an order of magnitude
larger than the ILC the electroweak effects are very considerably
enhanced at the LHC, being up to
$30-40\%$.\cite{Maina:2004rb,Kuhn:2005gv} Among the issues still to be
understood in such calculations (related to the cancellation of real
and virtual corrections) is the question of whether experiments will
include events with $W$ and $Z$'s as part of their normal QCD event
sample, or whether instead such events will be treated separately.

\section{All-order calculations}

%
%
%
%
%

All-order calculations in QCD are based on the resummation of
logarithmically dominant contributions at each order. Such
calculations are necessary if one is to investigate the properties of
\emph{typical} events, for which each extra power of $\as$ is accompanied
by large soft and collinear logarithms.

The two main ways of obtaining all-order resummed predictions are
with exclusive Monte Carlo event generators, and with analytical
resummations. The former provide moderately accurate (leading log (LL)
and some parts of NLL) predictions very flexibly for a wide variety of
processes, with hadronisation models included. The latter are able to
provide the highest accuracy (at parton level), but usually need to be
carried out by hand (painfully) for each new observable and/or
process.  All-order calculations are also used in small-$x$ and
saturation physics, which will be discussed briefly at the end of this
section.

\subsection{Monte Carlo event generators (MC)}

Various issues are present in current work on event generators --- the
switch from Fortran to C++; improvements in the showering algorithms
and the modelling of the underlying event; and the inclusion of
information from fixed-order calculations.

The motivation for moving to C++ is the need for a more modern and
structured programming language than Fortran 77.  C++ is then a
natural choice, in view both of its flexibility and its widespread use
in the experimental community.

Originally~it~was intended that Herwig\cite{Corcella:2000bw},
Pythia\cite{Sjostrand:2003wg} and
Ariadne/LDC\cite{Lonnblad:1992tz,Kharraziha:1997dn} should all make
use of a general C++ event generator framework known as
ThePEG.\cite{ThePEG} Herwig++,\cite{Gieseke:2003hm} based on ThePEG,
was recently released for $e^+e^-$ and work is in progress for a
hadron-hadron version. Pythia~7 was supposed to have been
the Pythia successor based on the ThePEG, however instead a standalone
C++ generator Pythia~8 is now being developed,\cite{Pythia8} perhaps
to be interfaced to ThePEG later on.  Another independent C++ event
generator has also recently become available, SHERPA,\cite{SHERPA}
whose showering and hadronisation algorithms are largely based on
those of Pythia, and which is already functional for hadron
collisions.

In both the Pythia and Herwig `camps' there have been developments on
new showering algorithms. Herwig++ incorporates an improved
angular-ordered shower\cite{Gieseke:2003rz} in which the `unpopulated'
phase space regions have been shrunk.  Pythia~6.3 has a new parton
shower\cite{Sjostrand:2004ef} based on transverse momentum ordering
(\ie somewhat like Ariadne) which provides an improved description of
$e^+e^-$ event-shape data and facilitates the modelling of multiple
interactions in hadron collisions. Separately, investigations of
alternatives to standard leading-log backward evolution algorithms for
initial-state showers are also being
pursued.\cite{Tanaka:2003gk,Jadach:2005yq}

\subsection{Matching MC \& fixed order}

Event generators reproduce the emission patterns for soft and
collinear gluons and also incorporate good models of the transition to
hadrons.  They are less able to deal with multiple hard emissions,
which, as discussed above, are important in many new particle
searches. There is therefore a need to combine event generators with
fixed-order calculations.
%

The main approach for this is the CKKW\cite{CKKW} proposal. Events are
generated based on the $n$-parton tree matrix elements (for various
$n$), keeping an event only if its $n$ partons are sufficiently well
separated to be considered as individual jets (according to some
threshold jet-distance measure, based \eg on relative $k_t$). Each
event is then assigned a `best-guess' branching history, using which
it can be reweighted with appropriate Sudakov form factors (to provide
virtual corrections) and running couplings. The normal parton
showering is then added on to each event at scales below the threshold
jet-distance measure. Over the past two years this method has become
widely adopted and is available in all major
generators.\cite{Mrenna:2003if}

As well as seeking to describe more jets it is also important to
increase the accuracy of event generators for limited number of jets,
by including NLO corrections. Here too there is one method that has so
far dominated practical uses, known as MC@NLO.\cite{Frixione:2002ik}
Very roughly, it takes the standard MC and modifies it according to
\begin{equation}
  \label{eq:MC@NLO}
  \mathrm{MC} \to \mathrm{MC} ( 1 + \mathrm{NLO} -
  \mathrm{NLO}_{\mathrm{MC}})\,,
\end{equation}
where $\mathrm{NLO}_{\mathrm{MC}}$ represents the effective NLO
corrections present by default in the MC. Since the MC usually has the
correct soft and collinear divergences, the combination $\mathrm{NLO}
- \mathrm{NLO}_{\mathrm{MC}}$ should be finite. Eq.~(\ref{eq:MC@NLO}) is
therefore a well-behaved way of introducing exactly the correction
needed to guarantee NLO correctness. A prediction from MC@NLO for
$b$-production,\cite{Frixione:2003ei} figure~\ref{fig:b}, is compared
to data and a purely analytical approach,\cite{Cacciari:2003uh} and
one notes the good agreement between all three.

\begin{figure}[htbp]
  \centering
  \includegraphics[width=\columnwidth]{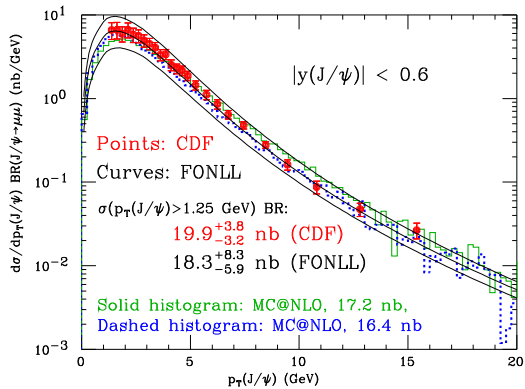}
  \caption{Spectrum for $b$-production ($\to J/\Psi$) compared to
    MC@NLO and an analytical prediction.\protect\cite{Cacciari:2003uh}}
  \label{fig:b}
\end{figure}

A bottleneck in widespread implementation of the MC@NLO approach is
the need to know $\mathrm{NLO}_{\mathrm{MC}}$, which is different for
each generator, and even process. Furthermore MC@NLO so far
guarantees NLO correctness only for a fixed number of jets --- \eg it
can provide NLO corrections to $W$ production, but then $W+1\jet$ is
only provided to LO. An approach to alleviate both these problems
proposes\cite{Nagy:2005aa} to combine, for example, $W$, $W+1\jet$,
$W+2\jets$, etc.\ with a procedure akin to CKKW. It seeks to alleviate
the problem of needing to calculate $\mathrm{NLO}_{\mathrm{MC}}$ as
follows: when considering $W+m\jets$ the $(m+1)^\mathrm{th}$ emission
(that needed for NLO accuracy) is generated not by the main MC, but by
a separate mini well-controlled generator, designed specifically for
that purpose and whose NLO expansion is easily calculated (as is then
the analogue of eq.(\ref{eq:MC@NLO})). The only implementation of this
so far (actually of an earlier, related formalism\cite{Kramer:2003jk})
has been for $e^+e^-$.\cite{Kramer:2005hw}

\subsection{Analytical resummations}

It is in the context of analytical resummations that one can envisage
the highest resummation accuracies, as well as the simplest matching
to fixed order calculations. Rather than directly calculating the
distribution $d\sigma(V)/dV$ of an observable $V$, one often considers
some integral transform $F(\nu)$ of the distribution so as to reduce
$F(\nu)$ to the form
\begin{equation}
  \label{eq:resummed}
  \ln F(\nu) = \sum_n (\underbrace{\as^n L^{n+1}}_\mathrm{LL} +
  \underbrace{\as^n L^n}_\mathrm{NLL} + \ldots)\,,
\end{equation}
with $L = \ln \nu$.

Much of the information for certain N$^3$LL threshold resummations was
recently provided by the MVV group.\cite{Moch:2005ba} The highest
accuracy for full phenomenological distributions is for the recently
calculated Higgs transverse momentum spectrum\cite{Bozzi:2005wk} at
the LHC and the related\cite{Dokshitzer:1978hw}
energy-energy-correlation (EEC) in $e^+e^-$,\cite{deFlorian:2004mp}
both of which have also been matched to NLO fixed order.

Boson transverse momentum spectra and the EEC are among the simplest
observables to resum. For more general observables and processes, such
as event shapes or multi-jet events, the highest accuracy obtained so
far has been NLL, and the calculations are both tedious and
error-prone. This has prompted work on understanding general features
of resummation. One line of research\cite{Bonciani:2003nt} examines
the problem of so-called `factorisable' observables, for an arbitrary
process. This extends the understanding of large-angle soft colour
evolution logarithms, whose resummation was originally pioneered for
4-jet processes by the Stony Brook
group,\cite{SB-SoftColour,SB-RapGap} in which unexplained hidden
symmetries were recently
discovered,\cite{Dokshitzer:2005ig,Seymour:2005ze} notably between
kinematic and colour variables.

Separately, the question of how to treat general observables has led
to a procedure for automating resummations for a large class of
event-shape-like observables.\cite{Banfi:2004yd} It avoids the need to
find an integral transform that factorises the observable and
introduces a new concept, recursive infrared and collinear safety,
which is a sufficient condition for the exponentiated form
eq.~(\ref{eq:resummed}) to hold. Its main application so far has been
to hadron-collider dijet event shapes,\cite{Banfi:2004nk} (see
also\cite{Sterman:2005bf}) which provide opportunities for
experimental investigation of soft-colour evolution and of
hadronisation and the underlying event at the Tevatron and LHC.

The above resummations apply to `global' observables, those sensitive
to radiation everywhere in an event. For non-global observables, such
as gap probabilities\cite{NGL-gap} or properties of individual
jets,\cite{NGL-jet,BD-dijet} a new class of enhanced term appears,
non-global logarithms $\as^n L^n$ (NGL), fig.~\ref{fig:NG-example}.
Their resummation has so far only been possible in the large-$N_C$
limit,\cite{NGL-jet,Banfi:2002hw} though the observation of structure
related to BFKL evolution,\cite{Marchesini:2003nh} has inspired
proposals for going to finite $N_C$.\cite{Weigert:2003mm}

\begin{figure}[htbp]
  \centering
  \includegraphics[width=0.6\columnwidth]{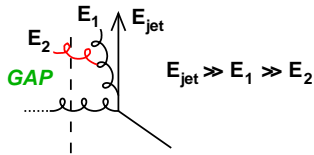}
  \caption{Diagram giving NGL: to calculate the probability of there
    being no emission into the gap, one should resum a large-angle
    energy-ordered cascade of emissions rather than just direct
    emission from the original hard partons.}
  \label{fig:NG-example}
\end{figure}

Non-global and soft colour resummations (in non-inclusive form\cite{SB-RapGap})
come together when calculating the probability of a gap between a pair
of jets at the Tevatron/LHC, relevant as a background to the $W$
fusion process for Higgs production.\cite{Appleby:2003sj} Recently it
has been pointed out that contributions that are subleading in
$1/N_C^2$ play an important role for large gaps,\cite{Forshaw:2005sx}
and also that there are considerable subtleties when using a $k_t$ jet
algorithm to to define the gap.\cite{Appleby:2002ke}

\subsection{Small $x$ and saturation}

The rise of the gluon at small $x$, as predicted by BFKL,\cite{BFKL}
leads eventually to such high gluon densities that a `saturation'
phenomenon should at some point set in. Usually one discusses this in
terms of a saturation scale, $Q^2_s(x)$, below (above) which the gluon
distribution is (un)saturated. From HERA data, it is believed that at
$x = x_0 \simeq 10^{-4} - 10^{-5}$, $Q_s$ is of order $1\GeV$ and that
it grows as $Q^2_s(x) = (x/x_0)^{-\lambda}\GeV^2$ with $\lambda\simeq
0.3$.\cite{Bartels:2002cj} $Q_s$ may be of relevance to the LHC
because the typical transverse scale $E_{m}$ of minimum bias minijets
should satisfy the relation $Q_s^2(s/E_{m}^2) \sim E_{m}^2$, whose
solution gives $E_m \sim (s x_0)^{\frac{\lambda}{2+\lambda}}
\GeV^{\frac{2-\lambda}{2+\lambda}} \simeq 2.7 - 3.6 \GeV$, or if one
doesn't trust the normalisation, a factor 1.7 relative to the
Tevatron. The exact phenomenology is however
delicate.\cite{Khoze:2004hx,Thorne:2005kj}

The theoretical study of the saturation scale has seen intense
activity these past 18 months, spurred by two observations. Firstly it
was pointed out\cite{Munier:2003sj} that the Balitsky-Kovchegov (BK)
equation, often used to describe the onset of saturation and the evolution
$d\ln Q^2_s/d\ln x$, is in the same universality class as the Fisher
Kolmogorov Petrovsky Piscounov (FKPP) reaction-diffusion equation,
much studied in statistical physics\cite{EbertVanSaarloos} and whose
travelling wave solutions relate to the evolution of the saturation
scale. Secondly, large corrections were
discovered\cite{Mueller:2004se} when going beyond the BK mean-field
approximation: to LO, $\lambda_{BK} \propto \as$, while the non
mean-field corrections go as $\as/ \ln^2 \as^2$.  Such corrections
turn out to be a familiar phenomenon in stochastic versions of the
FKPP equation, with $\as^2$ in QCD playing the role of the minimum
particle density in reaction-diffusion systems with a finite number of
particles.\cite{Iancu:2004es} The stochastic corrections also lead to
a large event-by-event dispersion in the saturation scale.

These stochastic studies are mostly based on educated guesses as to
the form of the small-$x$ evolution beyond the mean-field
approximation. There has also been extensive work on finding the full
equation that replaces BK beyond the mean-field approximation, a
number of new formulations having been
proposed.\cite{Iancu:2004iy,Levin:2005au,Kovner:2005jc,Hatta:2005rn}
It will be interesting to examine how their solutions compare to the
statistical physics related approaches.

\section{Concluding remarks}

Of the topics covered here, the one that has been the liveliest in the
past year is that of `twistors'.\footnote{Second place goes to
  saturation.} Its full impact cannot yet be gauged, but the dynamic
interaction between the QCD and string-theory communities on this
subject will hopefully bring further important advances.

More generally one can ask if QCD is on track for the LHC.  Progress
over the past years, both calculationally and phenomenologically (\eg
PDF fitting) has been steady. Remaining difficulties, for example in
high (NNLO) and moderate (NLO) accuracy calculations are substantial,
however the considerable number of novel ideas currently being
discussed encourages one to believe that significant further advances
will have been made by the time LHC turns on.

\section*{Acknowledgements}

Many people have helped with the preparation of this talk, through
explanations of material I was unfamiliar with, suggestions of topics
to include, and useful comments. Among them: Z.~Bern, T.~Binoth,
J.~Butterworth, M.~Cacciari, M.~Ciafaloni, P.~Ciafaloni, D.~Comelli,
D.~Dunbar, Yu.~L.~Dokshitzer, R.~K.~Ellis, J.-P.~Giullet, D.~Kosower,
L.~Lonnblad, A.~H.~Mueller, G.~Marchesini, S.~Moretti, S.~Munier,
M.~H.~Seymour, A.~Vogt, B.~R.~Webber, G.~Zanderighi. I wish also to
thank the organisers of the conference for the invitation, financial
support, and warm hospitality while in Uppsala.

\small

\clearpage
\normalsize
\section*{DISCUSSION}

\begin{description}
\item[Bennie Ward] (Baylor University):\\
To what extent are the deduced MHV rules using twistors now proven?

\item[Gavin Salam{\rm :}] an outline of a field-theoretic proof for
  the MHV (CSW) rules was given in the same article\cite{BCFW} as the
  proof of the BCF rules. Recently (after the Lepton Photon Symposium)
  a more detailed version of the proof has
  appeared.\cite{Risager:2005vk}
 
\end{description}

\end{document}